\documentclass[aps,prb,twocolumn,groupedaddress,showpacs]{revtex4}
\usepackage{epsfig}
\usepackage[dvipsnames,usenames]{color}
\usepackage{amssymb}

\emergencystretch=\maxdimen
\begin{document}

\title{Charge Gaps at Fractional Fillings in Boson Hubbard Ladders}

\author{T. Ying,$^{1,2}$ G.G.~Batrouni,$^{3,4,5}$ G.X.~Tang,$^1$
  X.D.~Sun,$^1$ and R.T.~Scalettar$^2$}

\affiliation{$^1$Department of Physics, Harbin Institute of
Technology, Harbin 150001, China}

\affiliation{$^2$Physics Department, University of California, Davis,
California 95616, USA}

\affiliation{$^3$INLN, Universit\'e de Nice--Sophia Antipolis, CNRS,
1361 route des Lucioles, 06560 Valbonne, France}

\affiliation{ $^4$Institut Universitaire de France, 103, Boulevard
  Saint-Michel, 75005 Paris, France}

\affiliation{$^5$ Centre for Quantum Technologies, National
University of Singapore, 2 Science Drive 3, Singapore 117542}

\begin{abstract}
  The Bose-Hubbard Hamiltonian describes the competition between
  superfluidity and Mott insulating behavior at zero temperature and
  commensurate filling as the strength of the on-site repulsion is
  varied.  Gapped insulating phases also occur at non-integer
  densities as a consequence of longer ranged repulsive interactions.
  In this paper we explore the formation of gapped phases in coupled
  chains due instead to anisotropies $t_x \neq t_y$ in the bosonic
  hopping, extending the work of Crepin {\it et al.} [Phys. Rev. B 84, 054517 (2011)] on two coupled
  chains, where a gap was shown to occur at half filling for
  arbitrarily small interchain hopping $t_y$.  Our main result is
  that, unlike the two-leg chains, for three- and four-leg chains, a
  charge gap requires a finite nonzero critical $t_y$ to
  open. However, these finite values are surprisingly small, well
  below the analogous values required for a fermionic band gap to
  open.
\end{abstract}

\pacs{
71.10.Fd, % Lattice fermion models (Hubbard model, etc.)
%% 71.30.+h, % Metal-insulator transitions and other electronic transitions
02.70.Uu  % Applications of Monte Carlo methods
}

\maketitle

%%%%%%%%%%%%%%%%%%%%%%%%%%%%%%%%%%%%%%%%%%%%%%%%%%%%%%%%%%%%%%%%%%
\section{INTRODUCTION}
%%%%%%%%%%%%%%%%%%%%%%%%%%%%%%%%%%%%%%%%%%%%%%%%%%%%%%%%%%%%%%%%%%

Experiments on ultracold atomic gases have opened new possibilities in
the exploration of strongly correlated quantum physics\cite{jaksch98}.
Most significantly, the ratio of interaction to kinetic energy can be
readily tuned, something which is possible only with substantial effort
in condensed matter systems, e.g.~through the application of high
pressure in a diamond anvil cell.
The flexible nature of the optical lattices that can be generated using
interfering lasers has also allowed the study of different dimensionality
and dimensional crossover, as well as the systematic interpolation
between different geometries in a given dimension, e.g.~triangular to
kagome\cite{jo12} or square to triangular\cite{greif13}.

In the case of bosonic systems, a key focus has been on the superfluid
(SF) to Mott insulator (MI) quantum phase transition, which occurs at
zero temperature and commensurate filling as the ratio of kinetic
energy to on-site repulsion $U$ is tuned \cite{fisher89}.  The MI
phase is characterized by a nonzero charge gap $\Delta_c$: A plot of
density, $\rho$, versus chemical potential, $\mu$, exhibits plateaux
when $\rho$ takes on integer values.  The critical interaction
strength is now known to very high precision, e.g.~taking the value
$(t/U)_c = 0.05974(3)$ for a square lattice \cite{capogrosso08}.  With
longer range interactions, these plateaux can develop at non-integer
fillings as well.  For example, a sufficiently large near-neighbor
repulsion $V_1$ drives a checkerboard solid order at $\rho=\frac12$ on
a square lattice\cite{niyaz91}. Likewise, a next-near-neighbor
repulsion $V_2$ can cause stripe ordering at the same filling.  Both
patterns of charge ordering are accompanied by a non-zero gap.  Long
range (e.g.~dipolar) interactions give rise to a complex "devil's
staircase" structure in $\rho(\mu)$ \cite{capogrosso10,ohgoe12}.  The
number of plateaux which develop increases with system size,
indicating the presence of frustration \cite{ohgoe12}.
  Another fascinating, and experimentally realizable, mechanism which
  produces MI phases at fractional fillings is the presence of a
  superlattice superimposed on the optical lattice. These superlattice
  fractional filling MI phases are expected in
  one\cite{buonsante04,buonsante05,rousseau06} and
  two\cite{buonsante05b,santos04} dimensions and in one-dimensional
  chains of tilted double well
  potentials\cite{danshita07,danshita08}.

In this paper we study the appearance of fractional filling MI
plateaux in the Bose-Hubbard model on coupled chains which
originate instead due only to {\it anisotropic hopping},
\begin{eqnarray}
\label{Hamiltonian}
&&{\hat\mathcal H}=
- \mu \sum_{i}  n_{i}
+ \frac12 U \sum_{i} n_{i} (n_{i} -1 )
\\
&&- t_x \sum_{ i }
(a_{i}^\dagger a_{i+\hat x}^{\phantom{\dagger}} +
a_{i+\hat x}^\dagger a_{i}^{\phantom{\dagger}} )
- t_y \sum_{ i }
(a_{i}^\dagger a_{i+\hat y}^{\phantom{\dagger}} +
a_{i+\hat y}^\dagger a_{i}^{\phantom{\dagger}} )
\nonumber
\label{ham}
\end{eqnarray}
Here $a_{i}^\dagger$, $a_{i}^{\phantom{\dagger}}$, and
$n_{i}^{\phantom{\dagger}}$ are boson creation, destruction, and
number operators on sites $i$ of an $L_x\times L_y$ rectangular
lattice.  $t_{x}$ and $t_{y}$ are hopping parameters between sites
${i}$ and neighoring sites in the $\hat x$ and $\hat y$ directions
respectively.  We will focus on "coupled chain" geometries in which
$L_x \gg L_y$.  $U$ is an on-site repulsion, We shall consider here
only the hard-core limit $U=\infty$. We note the absence of longer
range interactions and superlattice potentials, so that the only
possible mechanism for the fractional filling MI is the new one
(anisotropic hopping) considered here.

As noted above, the presence of density plateaux in the ground state
of the Hamiltonian, Eq.~(1), is expected at commensurate filling for
sufficiently large $U$, but otherwise a compressible SF phase is the
most natural low temperature phase.  In contrast, for fermions,
additional band insulator plateaux can arise from the structure of the
dispersion relation $\epsilon({\bf k})$.  For example, for a two-chain
($L_y=2$) system $\epsilon(k_x,k_y) = -2t_x \,{\rm cos}k_x -t_y \,{\rm
  cos}k_y$ where $k_y$ takes on only the two values $k_y=0,\pi$.  This
dispersion relation can equivalently be viewed as that of a single one-dimensional (1D)
chain with two bands, $\epsilon(k_x) = -2t_x \,{\rm cos}k_x \pm t_y$.
If $t_y > 2 t_x$ the two bands are separated by a gap, and a plateau
in $\rho(\mu)$ will occur at $\rho=\frac12$.

Such gaps arising from the structure of $\epsilon({\bf k})$ are
typically to be expected only of fermionic systems, since they occur
as a consequence of the Pauli principle and the complete filling of a
lower energy band.  In one dimension, however, hard-core ($U=\infty$)
bosons (the "Tonks-Girardeau gas")\cite{girardeau60} share many
similarities with fermions, via a Jordan-Wigner
transformation\cite{jordan28}.  Hence such band insulating behavior
might be expected for hard-core bosons as well.  Indeed, such 1D
"bosonic band insulators" have been observed when a superlattice
potential $\Delta \sum_i (-1)^i n_i$ is added to the 1D hard core
boson Hubbard Hamiltonian \cite{rousseau06}.  As in the two-chain
case, a superlattice potential opens a gap in the dispersion relation
which can be measured by the plateau in $\rho(\mu)$ and which is
accompanied by a vanishing of the SF density $\rho_s$.  Interestingly,
these insulators are found to extend deep into the soft-core limit,
where the Jordan-Wigner mapping to fermions is no longer valid.

Crepin {\it et al.$\,$}\cite{crepin11} have recently studied another
example of the persistence of the 1D Jordan-Wigner analogy between
hard-core bosons and fermions by examining two-leg ladders.  Here the
possibility of particle exchange destroys the formal mapping and one
might expect features of the fermionic case not to have bosonic
analogs.  Nevertheless, they found that in the large $t_y$ limit the
bosonic system exhibits a charge gap $\Delta = 2t_y - 4 t_x +
2t_x^2/t_y$, where the first two terms are precisely the fermionic
result.  The last term represents an {\it increase} in the gap for
hard-core bosons relative to fermions.  In addition, for weakly
coupled chains where $t_y \ll t_x$, the fermionic case would have
overlapping bands and no charge gap.  Yet for bosons $\Delta$ remains
non-zero, although exponentially small, with $\Delta \propto
e^{-\alpha t_x/t_y}$.

It is natural to consider the extension of the question of insulating
behavior in bosonic systems to cases where there are more than two
chains, and even to the two-dimensional limit. In this paper we present results for
the charge gap and SF response of the Hamiltonian, Eq.~(1), on three- and
four-leg ladders.  Our primary methodology is quantum Monte
Carlo (QMC) simulations using the ALPS\cite{alps11} stochastic series
expansion (SSE) code.  We supplement this with density matrix
renormalization group (DMRG) calculations also using the ALPS library.
Our major result is that incommensurate gaps persist beyond the two
leg case, although a finite, but surprisingly small, $t_y$ seems to be
required.  For the three- and four-chain systems we find that even for
$t_y/t_x = {\cal O}(1) $, gaps form at fractional fillings. As the
number of chains increases, the finite critical value
increases.

\begin{figure}[htp]
  \epsfig{figure=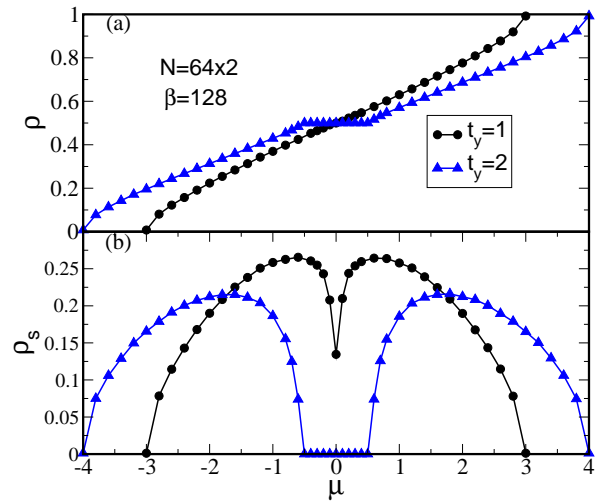,width=9.5cm,clip}
  \caption{(Color online) (a) $\rho(\mu)$ and (b) SF density for a $64
    \times 2$ ladder of hard-core bosons with $t_x=t_y=1$ (black) and
    $t_x=1$, $t_y=2$ (blue). For $t_y=2$, a gap is visible at
    $\rho=\frac12$, and the corresponding SF density vanishes,
    signaling the insulating state. At $t_y=1$ the gap is not visible
    in the $\rho(\mu)$ plot but the dip in $\rho_s(\mu=0)$ hints at
    its presence. Careful finite size scaling \cite{crepin11}
    demonstrates this clearly. Here and in subsequent figures
    error bars are smaller than the symbol size and hence are not shown.  }
\label{fig:muandrhos-64x2tx1ty1to2V0-SSE}
\end{figure}

%%%%%%%%%%%%%%%%%%%%%%%%%%%%%%%%%%%%%%%%%%%%%%%%%%%%%%%%%%%%%%%%%%
\section{Methodology}
%%%%%%%%%%%%%%%%%%%%%%%%%%%%%%%%%%%%%%%%%%%%%%%%%%%%%%%%%%%%%%%%%%

\vskip0.1in

Our simulations are performed using the SSE
algorithm\cite{sandvik99,alet05,pollet05}, a powerful and elegant QMC
method to study quantum spin or bosonic lattice models.  SSE is a
generalization of Handscomb's algorithm \cite{handscomb62} for the
Heisenberg model. It starts from a Taylor expansion of the partition
function in orders of inverse temperature $\beta$, and corresponds to
a perturbation expansion in all terms of the Hamiltonian.  There are
no ``Trotter errors" associated with discretization of imaginary time.
We use a grand-canonical formulation of the code, but also restrict
some measurements to just one particle number sector in order to
generate results in the canonical ensemble.

To characterize the phases of the Hamiltonian, Eq.~(1), we
will examine the energy, $E=\langle H \rangle$, the variation of
density as a function of chemical potential and the SF density
$\rho_s$. The SSE code computes the SF density via the
relation\cite{pollock}
\begin{equation}
\rho_s=\frac{\langle W^2\rangle}{2td\beta L^{d-2}},
\label{rhos}
\end{equation}
where $W$ is the winding number of the particle world lines, $d$ is
the dimensionality and $\beta$ is the inverse temperature. In the
coupled-chain problem we address here, we take $L_y\ll L_x$, $t_x\neq
t_y$, and the boundary conditions are open (periodic) in the $\hat y$ ($\hat x$)
direction. Note, however, that taking periodic boundary
  conditions in the $\hat y$ direction does not change the physics
  qualitatively; the fractional filling gaps are still present but the
  values of the critical $t_y$ change slightly. We focus, therefore,
on the SF density in the $\hat x$ direction so that the hopping parameter
in Eq.~(\ref{rhos}) is $t_x$ and the length $L$ is $L_x$. In what
follows, we typically take $\beta=2L_x$ to access ground state
properties. However, for small $L_x$ where $2L_x \leq 128$, we set
$\beta=128$. In addition, for some cases of $2L_x> 128$, we verified
that putting $\beta=3L_x$ yields the same results as $\beta=2L_x$. We
also take $t_x=1$ to set the energy scale.

%%%%%%%%%%%%%%%%%%%%%%%%%%%%%%%%%%%%%%%%%%%%%%%%%%%%%%%%%%%%%%%%%%
\section{Results}
%%%%%%%%%%%%%%%%%%%%%%%%%%%%%%%%%%%%%%%%%%%%%%%%%%%%%%%%%%%%%%%%%%

\begin{figure}[htp]
\epsfig{figure=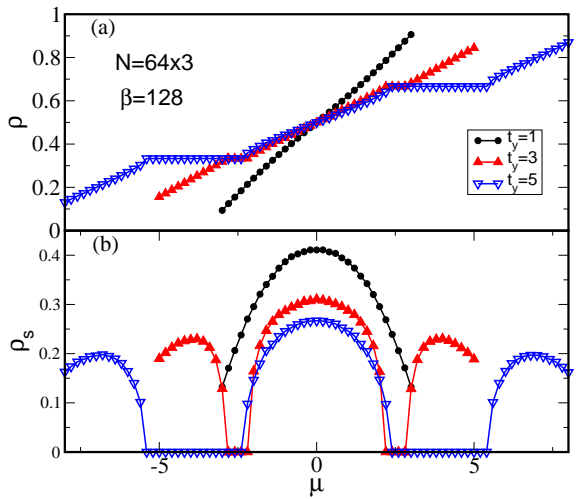,width=9.5cm,clip}
\caption{(Color online) (a) $\rho(\mu)$ and (b) SF density for a
  three-leg ladder of hard-core bosons.  As the transverse hopping
  $t_y$ grows, gaps develop at $\rho=\frac13, \frac23$.  When the
  chemical potential lies in the insulating gap, the SF density
  vanishes.}
\label{fig:muandrhos-64x3b128tx1ty1to6V0-SSE}
\end{figure}

We begin by studying the two-chain case considered by Crepin {\it et
  al.}$\,$\cite{crepin11} in order to confirm our methodology.  Figure
\ref{fig:muandrhos-64x2tx1ty1to2V0-SSE} demonstrates that our SSE
results are fully consistent with this earlier work.  A plateau in the
density at half filling is clearly visible for $t_y=2$ when $-0.5
\lesssim \mu/t \lesssim 0.5$, signaling the incompressible insulating
state where the SF density vanishes.  For $t_y=t_x=1$, the plateau at
half filling is not easily visible, but a sharp minimum appears in the
SF density hinting at its presence. A careful finite size scaling
study \cite{crepin11} shows its presence clearly.

We now explore whether these plateaux persist in geometries having
more than two chains, and if so, at what fillings.  Figure
\ref{fig:muandrhos-64x3b128tx1ty1to6V0-SSE}(a) shows the density as a
function of chemical potential for a three-leg ladder of 64 sites in the $\hat x$ direction.
Plateaux are seen to develop at $\rho=\frac13, \frac23$ for
sufficiently large $t_y$.  Figure
\ref{fig:muandrhos-64x3b128tx1ty1to6V0-SSE}(b) gives the SF density,
and confirms that it vanishes in the gapped phase.  At first glance,
such incommensurate insulators might seem surprising for a model with
only on-site repulsion, since empty sites exist to which the bosons
can hop without large energy cost.

\begin{figure}[htp]
\epsfig{figure=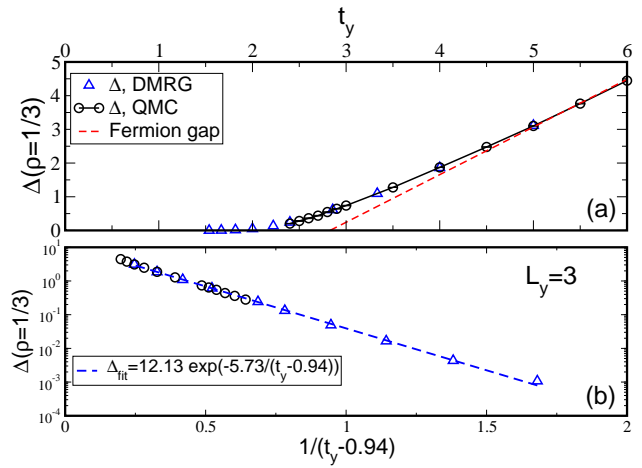,width=9.5cm,clip}
\caption{(Color online) (a) The gap $\Delta$ at $\rho = \frac13$ in a
  three-leg ladder as a function of $t_y$. The values of $\Delta$ are
  extrapolated from lattice sizes up to $L_x =250$.  For large $t_y$,
  $\Delta$ approaches the value appropriate for a collection of
  noninteracting fermions (dashed red line). (b) Semilogarithmic plot of the
  gaps in (a) as a function of $1/(t_y-0.94)$. }
\label{fig:3}
\end{figure}

As discussed already in the two-chain case, these insulators are not a
surprise if the particles are fermionic: For the noninteracting
system, $U=0$, the Hamiltonian can be diagonalized by going to
momentum space giving rise to the dispersion relation given in the
Introduction.  In the three leg case (with open boundary conditions in
the $\hat y$ direction), there are three bands, $\epsilon(k_x,k_y) = -2t_x
\,{\rm cos}k_x + \{-\sqrt{2}t_y, 0, \sqrt{2}t_y$\}.  For sufficiently
large $t_y$, a gap
$\Delta_f(\rho=\frac13)=0-2t_x-(-\sqrt{2}t_y+2t_x)=\sqrt2t_y-4$ is
present between the top of the lowest band and the bottom of the
middle band.  In a fermionic model at $T=0$ a band insulator would be
present at $\rho=\frac13$.  An identical gap would appear at
$\rho=\frac23$ as a consequence of particle-hole symmetry.  Thus
noninteracting fermions hopping on this three chain geometry would be
band insulators at $\rho=\frac13, \frac23$ for $t_y \geq 4/\sqrt{2}
\sim 2.829$ (see Fig.~\ref{fig:3}). The gaps at large
  values of $t_y/t_x$ are easy to understand in the hard core
  limit. When $t_x\ll t_y$, the particles delocalize much more
  efficiently along the $\hat y$ axis, and hence spread out in that
  direction.  Thus, when
  $\rho=1/3$, the particles are in an effective one-dimensional system
  with a density $\rho_{\rm eff}= 1$.  This system, due to the
  hard core nature of the particles, becomes an incompressible
  insulator. This intuitive picture applies equally to the four-chain
  system.  We examine below how small $t_y$ should be for such behavior
  to change.

The dependence of the bosonic gap on $t_y$ at $\rho=\frac13$ in a
three-leg system is shown in Fig.~\ref{fig:3}(a). The gap $\Delta(N)$ is computed
from the jump in the chemical potential $\mu(N)=E(N)-E(N-1)$ at the
critical density, i.e.~$\Delta(N)=\mu(N+1)-\mu(N)=E(N+1)+E(N-1)-2E(N)$.  We
use both the ALPS SSE and DMRG codes. The results are extrapolated
using lattice sizes up to $L_x = 250$ for the smaller values of the
gap.  In the DMRG calculations we keep $800$ states, which we have
verified to be converged.  The gap for a noninteracting three-leg
fermion chain is also presented.  The convergence of the bosonic
results to the fermionic ones at large $t_y$ suggests a fermion
mapping is appropriate even in the absence of a formal Jordan-Wigner
transformation.  The boson $\Delta$ is larger than its fermionic
counterpart and is unambiguously non-zero below the critical $t_y$
where fermions would become metallic.  A key issue is whether finite
$\Delta$ persists all the way down to $t_y=0$ as occurs for the
two-chain case \cite{crepin11}.

To try to answer this question, Fig.~\ref{fig:3}(b) shows a semilogarithmic
plot of the gap as a function of $1/(t_y-t_y^c)$ where $t_y^c$ is the
putative critical value at which the gap vanishes. A fit of the form
$\Delta=a\, {\rm exp}(-b/(t_y-t_y^c))$ yields $t_y^c \approx 0.94$ and
this is shown as the dashed (blue) line in the lower panel.  In the
two-chain case, Crepin {\it et al.} have found \cite{crepin11} an
exponential behavior all the way to very small (vanishing) $t_y$
(large $1/t_y$).  As a consequence, they argued that an exponentially
small gap persists all the way to $t_y=0$.  In the present three-leg case,
the evidence for a finite value for $t_y^c$ is clear.  The value is
considerably less than the fermionic band structure value $4/\sqrt{2}$
and even suggests that when the system is isotropic, $t_x=t_y$, the
three-leg system is gapped at $\rho =\frac13$ (and also $\rho =\frac23$).

\begin{figure}[htp]
\epsfig{figure=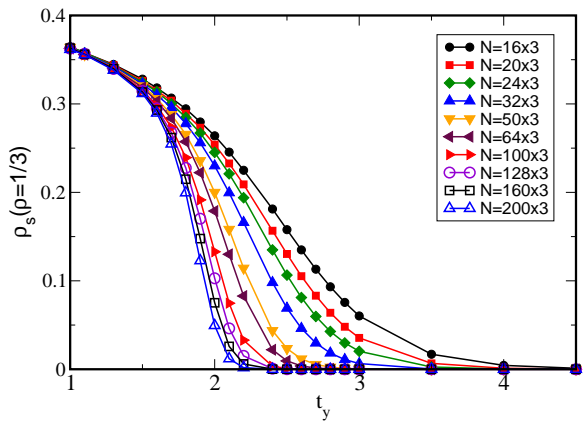,width=9.5cm,clip}
\caption{(Color online) SF density at $\rho =\frac13$ as a function of
  $t_y$.  The SF density remains nonzero for small $t_y$, even in the
  largest lattice size we study here.}
\label{fig:rhos-rho0333-3leg}
\end{figure}

\begin{figure}[htp]
\epsfig{figure=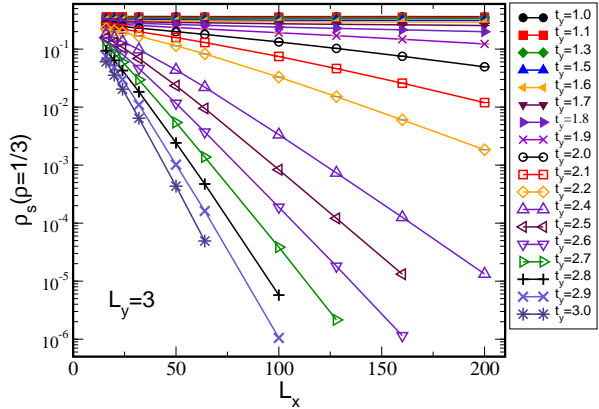,width=9.5cm,clip}
\caption{(Color online) Semilogarithmic plot of the SF density at $\rho
  =\frac13$ for different $t_y$, as a function of $L_x$.  In a non-SF
  phase, we expect $\rho_s \sim {\rm exp}(\,-L_x/\xi \,)$ and hence
  linear behavior here.  }
\label{fig:semilogrhos-rho0333-3leg}
\end{figure}

\begin{figure}[htp]
\epsfig{figure=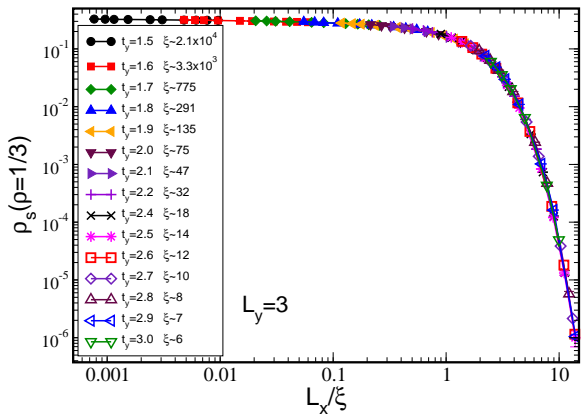,width=9.5cm,clip}
\caption{(Color online) Collapse of the data in
  Fig. \ref{fig:semilogrhos-rho0333-3leg}, by rescaling the $x$ axis,
  from $L_x \to L_x/\xi$, with the correlation length $\xi$ obtained
  from Fig. \ref{fig:semilogrhos-rho0333-3leg}.}
\label{fig:logrhos-rho0333-3leg}
\end{figure}

\begin{figure}[htp]
\epsfig{figure=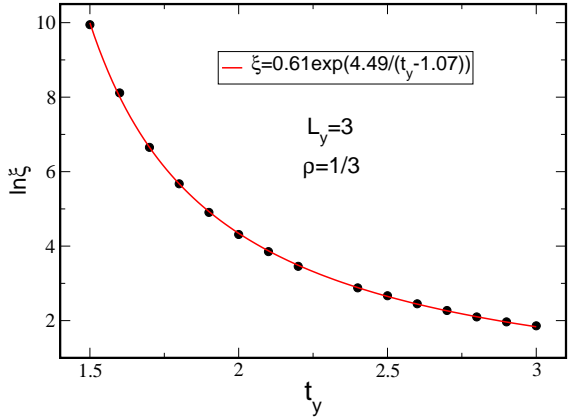,width=9.5cm,clip}
\caption{(Color online) Correlation length $\xi$ of the three-leg
  ladder versus the $\hat y$ direction hopping $t_y$. The points obtained
  from the data collapse (black circles) are well fitted by the
  exponential form (red curve).}
\label{fig:xi-3leg-rho0333}
\end{figure}

To confirm the above result for $t_y^c$, we turn to an examination of
the SF density at $\rho =\frac13$ to see if it becomes nonzero at
finite, small $t_y$.  Figure~\ref{fig:rhos-rho0333-3leg} shows
$\rho_s$ at $\rho =\frac13$ for lattice sizes $L_x = 16$ to $L_x =
200$. The SF density decreases with increasing $t_y$, and for $t_y
\gtrsim 4/\sqrt{2}$, the fermionic critical value for the three-band
case previously discussed, the behavior seems unambiguously insulating
($\rho_s=0$), consistent with the analysis of the charge gap.  Indeed
$\rho_s$ vanishes even for somewhat smaller $t_y$, in agreement with
the observation that the bosonic charge gap lies above the fermion
one.  However, for smaller $t_y$, the SF density is nonzero even for
the largest lattice size we consider here.

The persistence of nonzero $\rho_s$ as the lattice size increases in
Fig.~\ref{fig:rhos-rho0333-3leg} can be taken as an indication of a
nonzero $t_y^c$.  To verify this possibility quantitatively, we make a
semilogarithmic plot of the SF density versus $L_x$ for different $t_y$
(Fig.~\ref{fig:semilogrhos-rho0333-3leg}).  For large $t_y$, the SF
density decays exponentially with $L_x$ and we can compute correlation
length $\xi(t_y)$ from $\rho_s \simeq {\rm exp}(\,-L_x/\xi\,)$.  This
analysis parallels that presented in [\onlinecite{crepin11}]. The
values of $\xi$ are presented in the legend of
Fig.~\ref{fig:logrhos-rho0333-3leg}.  They are fairly well determined
down to $t_y \sim 1.7$ where $\xi$ is still on the order of the size
of our simulation box.  For smaller $t_y$ the correlation length
becomes much larger than $L_x$, as indicated by the nearly horizontal
traces in Fig.~\ref{fig:semilogrhos-rho0333-3leg}.  A consistency
check on our extraction of $\xi$ is provided by collapsing the data
for $\rho_s$, that is, via plotting results for different $t_y$ as a
function of a rescaled horizontal axis $L_x/\xi$ as in
Fig.~\ref{fig:logrhos-rho0333-3leg}.

\begin{figure}[htp]
\epsfig{figure=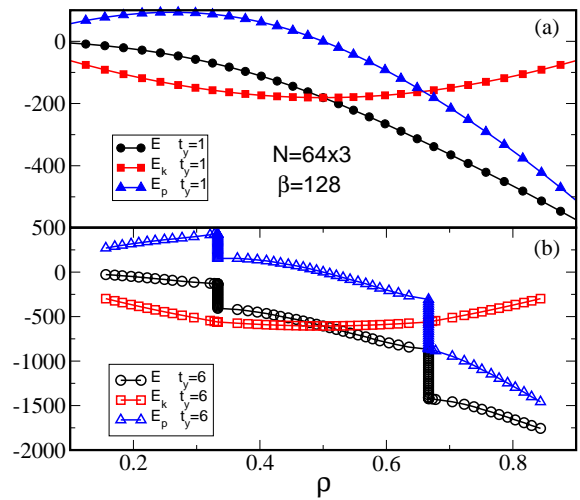,width=9.5cm,clip}
\caption{(Color online) Kinetic, potential, and total energies of the
  three-leg ladder as functions of the filling $\rho$ for (a) $t_y=1$
  and (b) $t_y=6$.  The kinetic energy is maximized (in absolute
  value) at half filling. While all the energy curves are smooth for
  $t_y=1$, for $t_y=6$ the curves of potential energy and total
  energy have vertical jumps at $\rho =\frac13$ and $\frac23$, also
  displayed by the kinetic energy although barely visible
    on the scale of the figure.  }
\label{fig:E-64x3b128tx1ty1and6V0}
\end{figure}

\begin{figure}[htp]
\epsfig{figure=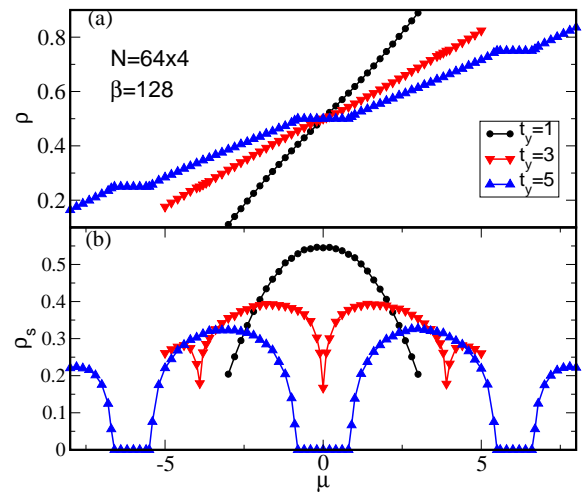,width=9.5cm,clip}
\caption{(Color online) (a) $\rho(\mu)$ and (b) SF density for a four
  leg ladder of hard-core bosons, similar to those in
  Fig.~\ref{fig:muandrhos-64x3b128tx1ty1to6V0-SSE}.  As the transverse
  hopping $t_y$ grows, gaps develop at $\rho=\frac14$, $\frac12$ and
  $\frac34$, and SF density vanishes there.  }
\label{fig:muandrhos-64x4b128tx1ty1to6V0-SSE}
\end{figure}

Figure~\ref{fig:xi-3leg-rho0333} provides a fit of these values of
$\xi$ to the functional form $\xi = \xi_0 \, {\rm
  exp}[\,a/(t_y-t_y^c)\,]$.  The best fit is obtained for $t_y^c
\approx 1.07$.  This value is in good agreement with the value,
$t_y^c\approx 0.94$, obtained above from a study of the charge gap and
confirms that a finite, but not very large, value of the transverse
hopping, $t_y$, is necessary to establish a gapped insulating phase at
$\rho =\frac13$.

In Fig.~\ref{fig:E-64x3b128tx1ty1and6V0}, we investigate the
individual components of the energy. The kinetic energy is maximized
(in absolute value) at half filling, because we are studying the case
with no intersite interactions. For $t_y=1$, all the energy curves are
smooth; for $t_y=6$, the curves of potential energy and total energy
have vertical jumps at $\rho =\frac13$ and $\frac23$, which
correspond to the insulating gap in
Fig.~\ref{fig:muandrhos-64x3b128tx1ty1to6V0-SSE}. The curve of the
kinetic energy also has kinks at those densities, although
these are barely visible at the scale of the figure.

\begin{figure}[htp]
\epsfig{figure=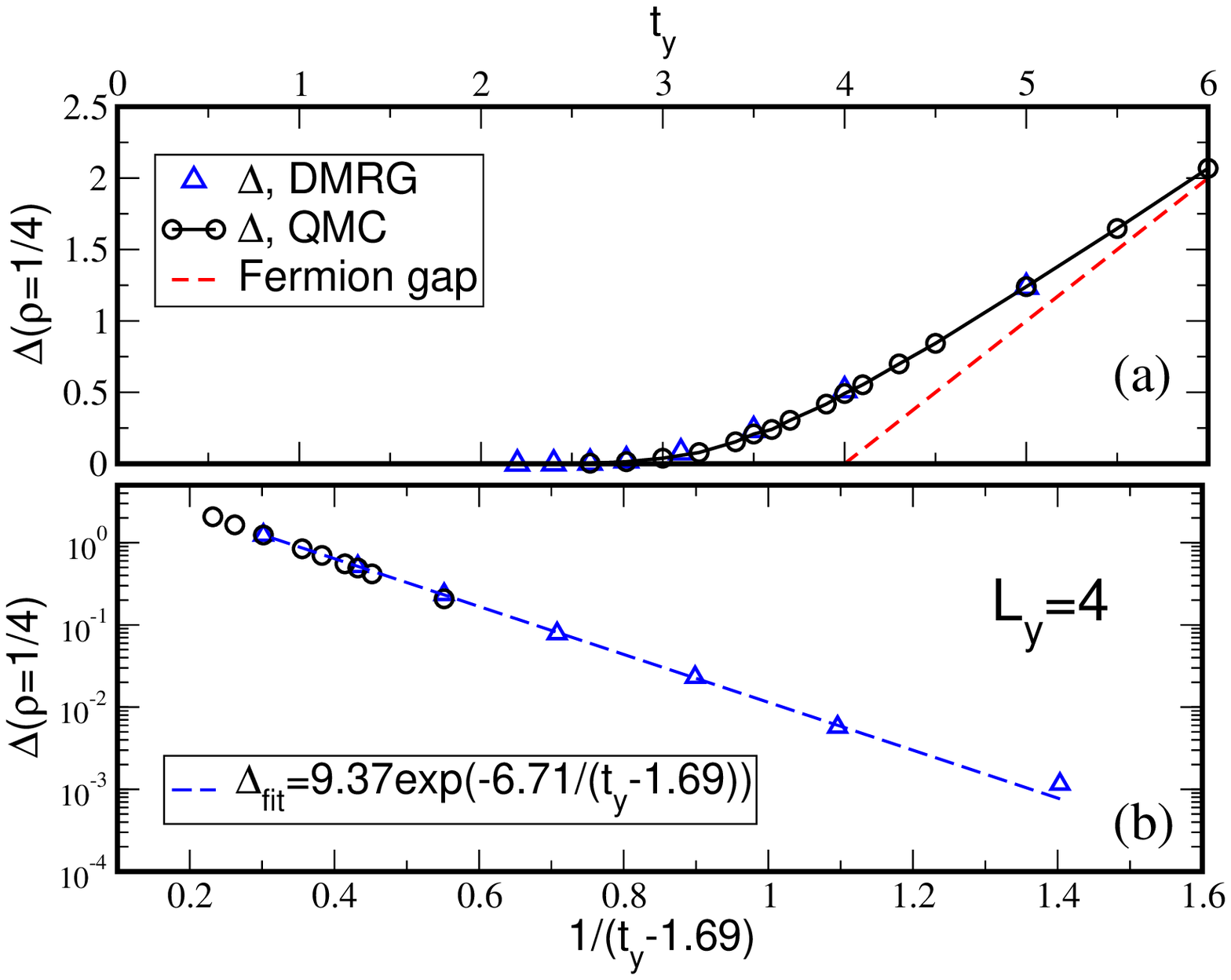,width=9.5cm,clip}
\caption{(Color online) (a) The extrapolated gaps at $\rho = \frac14$
  in the four-leg ladder as a function of $t_y$ from QMC and DMRG. (b)
  Semilogarithmic plot of the gap versus $(t_y-1.69)^{-1}$. Exponential
  decay over three decades is seen giving the critical value
  $t_y^c\approx 1.69$. }
\label{fig:4-rho025}
\end{figure}

\begin{figure}[htp]
\epsfig{figure=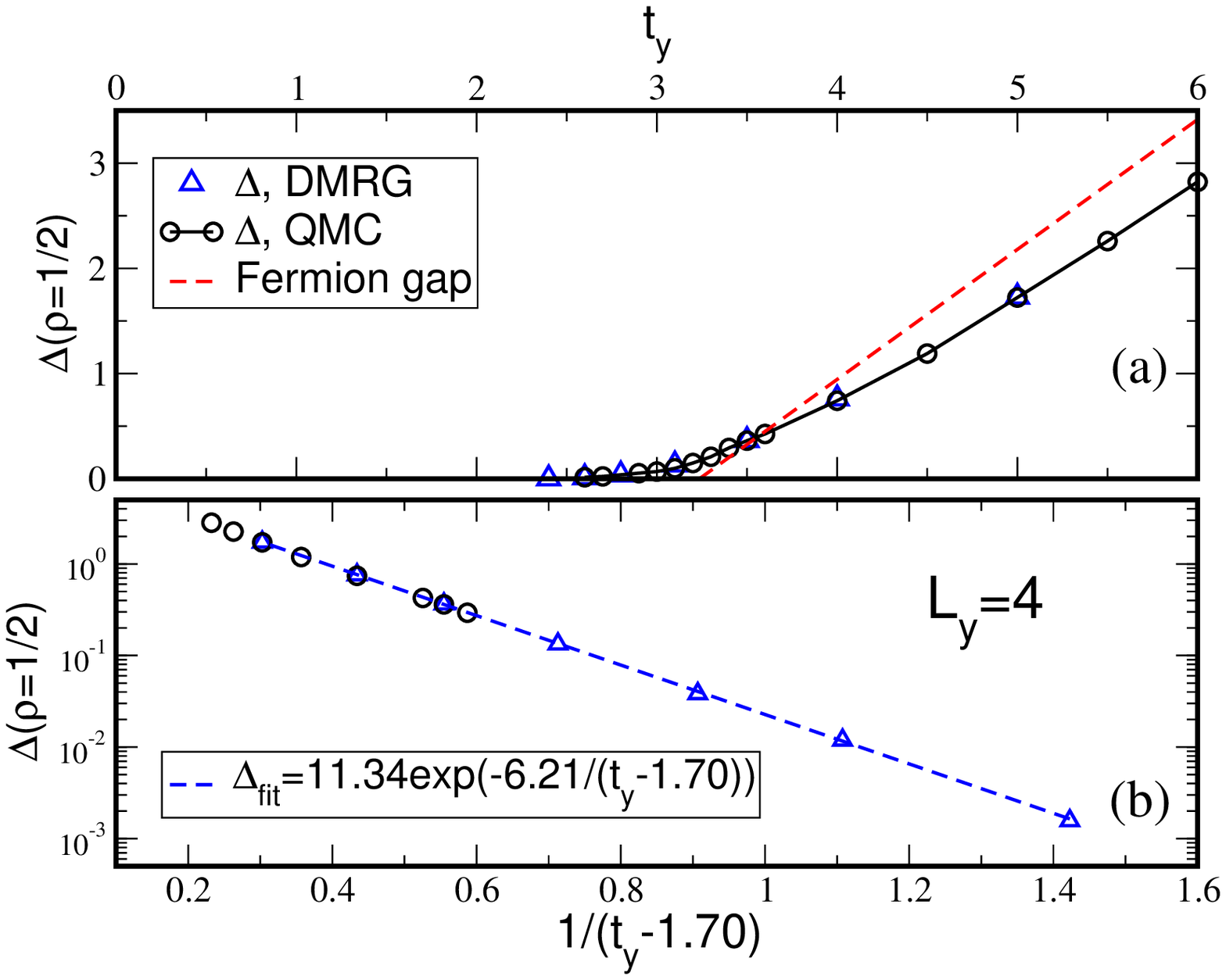,width=9.5cm,clip}
\caption{(Color online) (a) The extrapolated gaps at $\rho = \frac12$
  in the four-leg ladder as a function of $t_y$ from QMC and DMRG. (b)
  Semilogarithmic plot of the gap versus $(t_y-1.70)^{-1}$. Exponential
  decay over three decades is seen giving the critical value
  $t_y^c\approx 1.70$. }
\label{fig:4-rho05}
\end{figure}

Figure \ref{fig:muandrhos-64x4b128tx1ty1to6V0-SSE} is the same as Fig.
\ref{fig:muandrhos-64x3b128tx1ty1to6V0-SSE} but for the four-leg ladder
and shows (a) the density as a function of chemical potential for a
system of $64\times 4$ sites. Plateaux are seen to develop at $\rho
=\frac14, \frac12, \frac34$ for sufficiently large $t_y$. Figure
\ref{fig:muandrhos-64x4b128tx1ty1to6V0-SSE}(b) shows the SF density,
and confirms that it vanishes in the gapped phase. In this case the
fermionic band structure consists of four bands (two of which are
degenerate) with $\epsilon({\bf k}) = -2 t_x\,{\rm cos}k_x + 2 t_y \{
-1, 0, 0, +1 \}$, and the fermionic gaps
$\Delta_f(\rho=\frac14)=\Delta_f(\rho=\frac34)=t_y-4$,
$\Delta_f(\rho=\frac12)=\sqrt{2(3-\sqrt{5})}t_y-4$.  As opposed to
Fig.~\ref{fig:muandrhos-64x3b128tx1ty1to6V0-SSE} which showed two gaps
between the three bands, we have here three gaps at $\rho=\frac14,
\frac12, \frac34$ between the four bands.

Figures~\ref{fig:4-rho025} and \ref{fig:4-rho05} show in the (a)
panels gaps at $\rho =\frac14$ and $\rho =\frac12$ respectively as
functions of $t_y$ for the four-leg lattice, similar to
Fig.~\ref{fig:3}. As for the three-leg ladder, these gaps are obtained
with DMRG and QMC using the definition $\Delta(N)=E(N+1)+E(N-1)-2E(N)$
and then extrapolated from lattice sites $L_x = 16$ to $128$ (with
DMRG $L_x$ is taken up to $250$). The fermionic gaps are also included
in these figures. Compared with the three-leg ladder, the gaps in the
four-leg system need somewhat larger $t_y$ to form and, as expected,
the gaps at $\rho = \frac12$ are wider than at $\rho = \frac14$. The
fermionic band gaps provide a reasonable estimate of those seen in
the bosonic case.

%\begin{figure}[htp]
%\epsfig{figure=rhos-rho025-4leg.eps,width=9.5cm,clip}
%\caption{(Color online) SF density at $\rho =1/4$ as a function of
%  $t_y$.  The SF density remains nonzero for samll $t_y$, even in the
%  largest lattice size we study here.}
%\label{fig:rhos-rho025-4leg}
%\end{figure}

%\begin{figure}[htp]
%\epsfig{figure=semilogrhos-rho025-4leg.eps,width=9.5cm,clip}
%\caption{(Color online) Semi-log plot of the SF density at $\rho =1/4$
%  for different $t_y$, as a function of $L_x$. Exponential decays are
%  are clear for the larger $t_y$.}
%\label{fig:semilogrhos-rho025-4leg}
%\end{figure}

\begin{figure}[htp]
\epsfig{figure=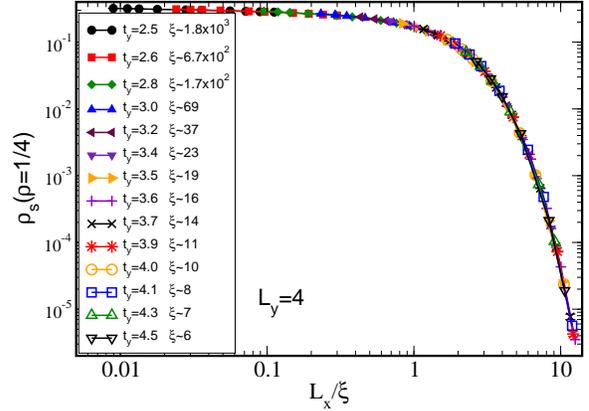,width=9.5cm,clip}
\caption{(Color online) The SF density, $\rho_s$, versus the
  scaled system size, $L/\xi$. $\xi$ is the correlation length at the
  values of $t_y$ shown in the legend. These values of $\xi$ are used
  in Fig. \ref{fig:xi-4leg-rho025} to obtain the critical value of the
  hopping, $t_y^c$.  }
\label{fig:logrhos-rho025-4leg}
\end{figure}

\begin{figure}[htp]
\epsfig{figure=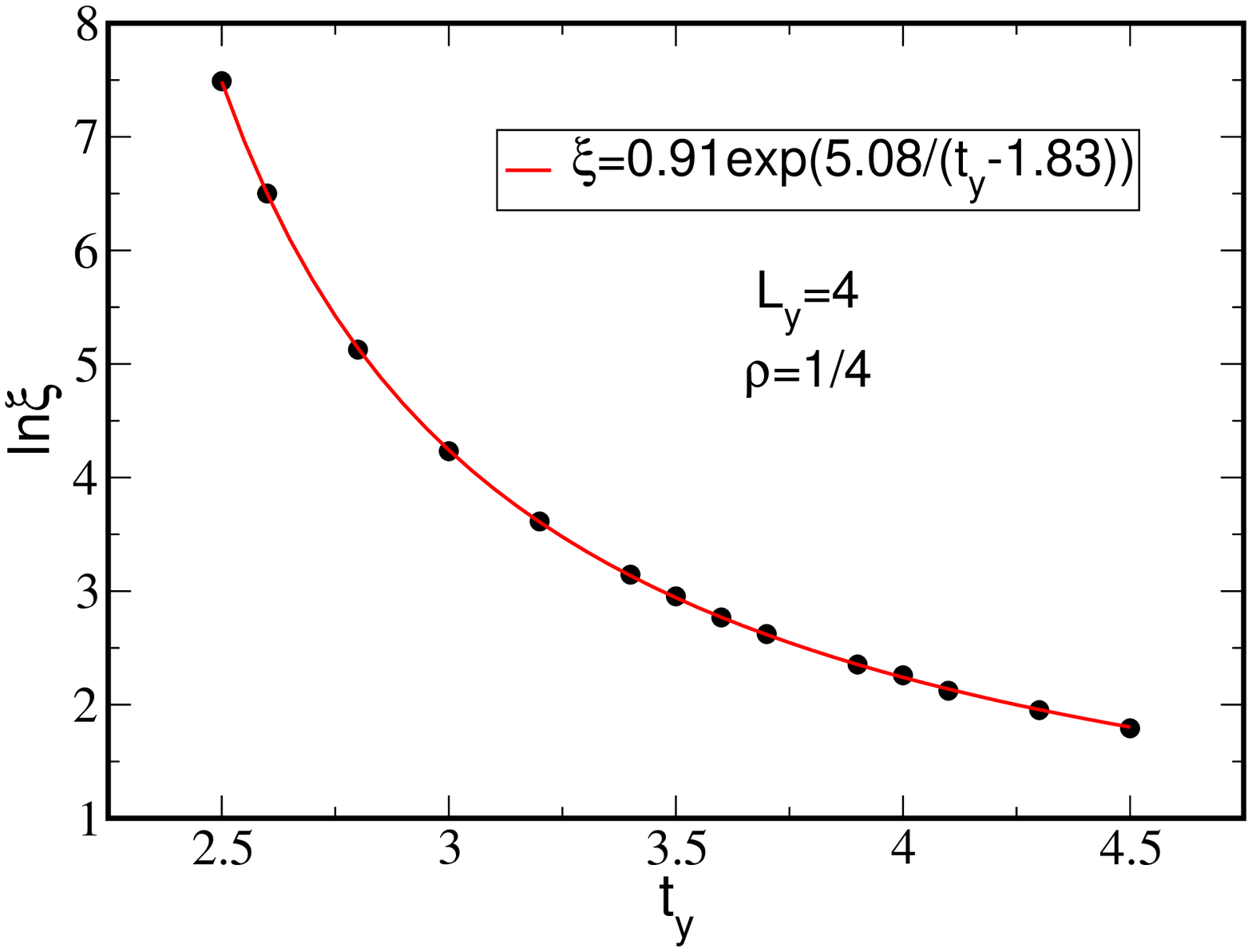,width=9.5cm,clip}
\caption{(Color online) Correlation length at $\rho=\frac14$ for the
  four-leg ladder versus the transverse hopping $t_y$. The points
  obtained from the data collapse (black circles) are fitted well by
  the exponential form (red curve) indicating a finite critical value
  $t_y^c\approx 1.83$.}
\label{fig:xi-4leg-rho025}
\end{figure}

%\begin{figure}[htp]
%\epsfig{figure=rhos-rho05-4leg.eps,width=9.5cm,clip}
%\caption{(Color online) SF density at $\rho =1/2$ as a function of
%  $t_y$.  The SF density remains nonzero for samll $t_y$, even in the
%  largest lattice size we study here.}
%\label{fig:rhos-rho05-4leg}
%\end{figure}

%\begin{figure}[htp]
%\epsfig{figure=semilogrhos-rho05-4leg.eps,width=9.5cm,clip}
%\caption{(Color online) Semi-log plot of the SF density at $\rho =1/2$
%  for different $t_y$, as functions of $L_x$. Exponential decays are
%  clear for the larger $t_y$.}
%\label{fig:semilogrhos-rho05-4leg}
%\end{figure}

\begin{figure}[htp]
\epsfig{figure=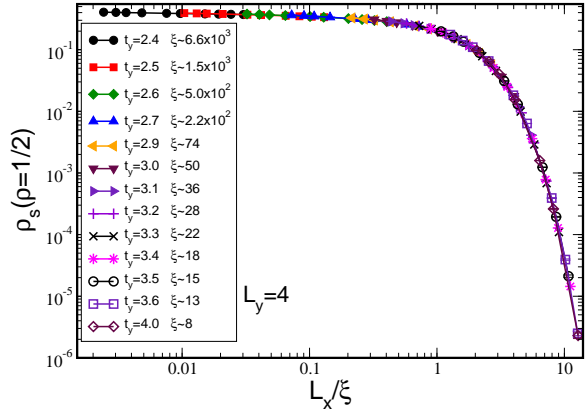,width=9.5cm,clip}
\caption{(Color online) As in Fig. \ref{fig:logrhos-rho025-4leg} but
  for $\rho =\frac12$. The values of $\xi$ are used in
  Fig. \ref{fig:xi-4leg-rho05} to obtain the critical value of the
  hopping, $t_y^c$. }
\label{fig:logrhos-rho05-4leg}
\end{figure}

\begin{figure}[htp]
\epsfig{figure=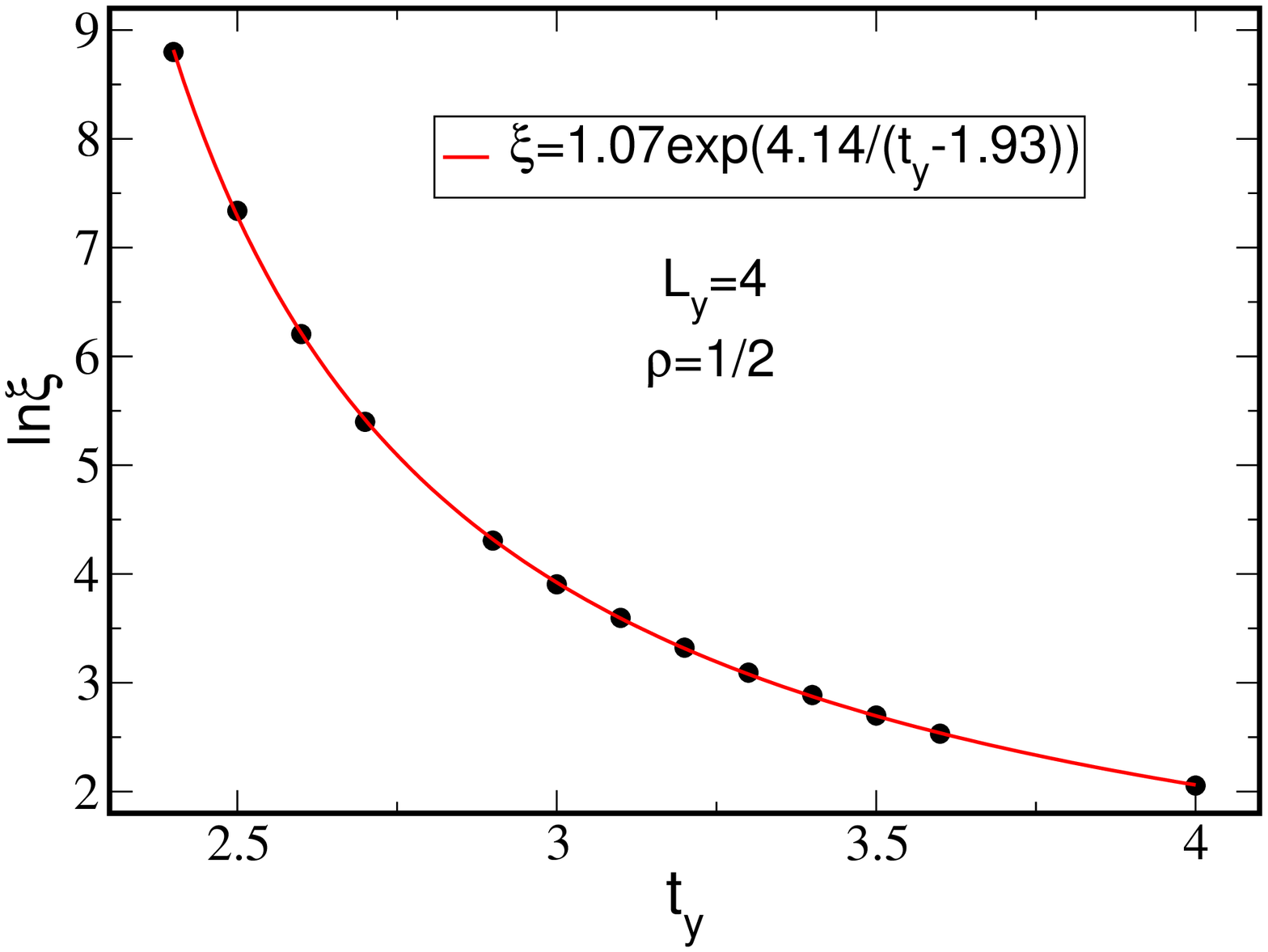,width=9.5cm,clip}
\caption{(Color online) Correlation length at $\rho =\frac12$ of the
  four-leg ladder versus the transverse hopping $t_y$. The points
  obtained from the data collapse (black circles) are fitted well by
  the exponential form (red curve). This gives the finite critical
  value for the hopping, $t_y^c\approx 1.93$.}
\label{fig:xi-4leg-rho05}
\end{figure}

We have also examined the exponential behavior of the gap at $\rho
=\frac14$ and $\rho =\frac12$ in the four-leg ladder as described for the
three-leg system. The (b) panels of Figs.~\ref{fig:4-rho025} and
\ref{fig:4-rho05} show semilogarithmic plots of the gaps as functions of
$(t_y-t_y^c)^{-1}$ with $t_y^c=1.69$ for $\rho =\frac14$ and
$t_y^c=1.70$ for $\rho =\frac12$. Exponential decay is seen over three
decades yielding finite values for the critical hoppings. As in the
three-leg case the critical values of $t_y$ at which gaps appear are
reduced from their fermion counterparts.

As for the three-leg ladder, we confirm the above results for
$t_y^c$ by examining the SF density at $\rho =\frac14$ and $\rho
=\frac34$ to see if it becomes nonzero at finite, small $t_y$.
The lattice size is up to $L_x=200$, as in the
three-leg case (Fig.~\ref{fig:rhos-rho0333-3leg}). We follow the
same analysis which led to Figs. \ref{fig:rhos-rho0333-3leg} and
\ref{fig:semilogrhos-rho0333-3leg} but only show figures for the
scaled SF density.  Figure \ref{fig:logrhos-rho025-4leg}
gives, for $\rho =\frac14$, $\rho_s$ versus the scaled size of the
system, $L_x/\xi$, where $\xi$ is the correlation length at the
various values of $t_y$ shown in the legend of the figure. The data
collapse is excellent and the values of $\xi$ thus obtained are shown
versus $t_y$ in Fig. \ref{fig:xi-4leg-rho025}. The exponential fit,
also shown in the figure, yields the critical value $t_y^c=1.83$, which
is in reasonable agreement with the value obtained from the gap,
$t_y^c=1.69$. Figures \ref{fig:logrhos-rho05-4leg} and
\ref{fig:xi-4leg-rho05} show the corresponding figures at $\rho
=\frac12$ and yield $t_y^c=1.93$ in good agreement with the value
obtained directly from the gaps, $t_y^c = 1.70$.

%%%%%%%%%%%%%%%%%%%%%%%%%%%%%%%%%%%%%%%%%%%%%%%%%%%%%%%%%%%%%%%%%%
\section{Conclusions}
%%%%%%%%%%%%%%%%%%%%%%%%%%%%%%%%%%%%%%%%%%%%%%%%%%%%%%%%%%%%%%%%%%

The most simple physical picture of the origin of Mott
insulating phases focuses on the (large) repulsive interaction
energy, and the filling at which the addition of quantum particles
first causes that energy to appear.  In a model with only a strong on-site
repulsion, particles can be added without any potential energy cost up
until a commensurate filling¡ª¡ªone particle per site.  At that point,
double occupancy becomes unavoidable, and the cost to increase the
density jumps.  The precise form of the kinetic energy, e.g., whether
it is isotropic in different spatial directions, or connects only
near-neighbor sites, is irrelevant to this simplistic argument:
As long as any empty sites remain in the lattice there is no charge
gap and, at low temperatures, a collection of bosonic particles
will form a SF phase. However, recent work by Crepin has
demonstrated that in two-leg ladders insulating phases can arise
at half-filling, underlining the need to refine this picture.

In this paper we studied hard-core bosons in three- and four-leg
ladders with anisotropic hopping parameters in the $\hat x$ and $\hat y$
directions, $t_x$ and $t_y$. To this end, we used both QMC and DMRG
from the ALPS library\cite{alps11}. Our focus was the possibility of
the appearance of gaps at fractional fillings as happens for the two-leg
ladder\cite{crepin11}, a possibility which seems surprising in light
of the argument above.  At the other extreme, when the number of
coupled chains is large and the system approaches the limit of a
two-dimensional system, one expects gaps to appear at fractional
fillings but at very large values of $t_y/t_x$\cite{ying13}.  We have
shown above that for both the three-leg and the four-leg systems gaps
appear at values of $t_y/t_x$, which are smaller than those resulting
from an argument based on fermionic bands. While in the case of the
two-leg system, the gap at $\rho =\frac12$ persists\cite{crepin11} all
the way down to $t_y=0$, we found that, in the three-leg case, a gap
appears at $\rho =\frac13$ starting at the isotropic value,
$t_y/t_x\approx 1$. For the four-leg system, a gap at $\rho =\frac14$
appears starting at $t_y/t_x \approx 1.8$ and for $\rho =\frac12$ at
$t_y/t_x\approx 1.9$.

\section*{ACKNOWLEDGMENTS}

This work was supported by: the National Key Basic Research Program of
China (Grant No.~2013CB328702), the National Natural Science Foundation of China (Grant
No. 11374074), a CNRS-UC Davis EPOCAL LIA joint research
grant, and by the University of California Office of the President.
We thank J. McCrea for useful input.

%%%%%%%%%%%%%%%%%%%%%%%%%%%%%%%%%%%%%%%%%%%%%%%%%%%%%%%%%%%%%%%%%%%%%%%%
%%%
%%%%%     BIBLIOGRAPHY
%%%%%%%%%%%%%%%%%%%%%%%%%%%%%%%%%%%%%%%%%%%%%%%%%%%%%%%%%%%%%%%%%%%%%%%%%%%


\begin{thebibliography}{100}

\bibitem{jaksch98}
%% ``Cold Bosonic Atoms in Optical Lattices,"
D. Jaksch, C. Bruder, J.I. Cirac, C.W. Gardiner, and P. Zoller,
Phys. Rev. Lett. {\bf 81}, 3108 (1998).

\bibitem{jo12}
G.-B. Jo, J. Guzman, C. K. Thomas, P. Hosur,
 A. Vishwanath, and D. M. Stamper-Kurn, Phys. Rev.
 Lett. {\bf 108}, 045305 (2012).

\bibitem{greif13}
D. Greif, T. Uehlinger, G. Jotzu, L. Tarruell, and
     T. Esslinger, Science {\bf 340}, 1307 (2013).

\bibitem{fisher89}
%% `Boson Localization and the superfluid--Insulator Transition,''
M.P.A. Fisher, P.B. Weichman, G. Grinstein
and D.S. Fisher, Phys. Rev. {\bf B 40}, 546 (1989).

\bibitem{capogrosso08}
B. Capogrosso-Sansone, S.G. S\"oyler, N. Prokof'ev,
and B. Svistunov,
Phys. Rev. {\bf A77}, 015602 (2008).

\bibitem{niyaz91}
P.~Niyaz, R.T.~Scalettar, C.Y.~Fong, and
G.G.~Batrouni, Phys.~Rev.~{\bf B 44}, 7143 (1991).

\bibitem{capogrosso10}
B. Capogrosso-Sansone, C. Trefzger, M. Lewenstein, P. Zoller, and
G. Pupillo,
Phys. Rev. Lett. {\bf 104}, 125301 (2010).

\bibitem{ohgoe12}
T. Ohgoe, T. Suzuki, and N. Kawashima,
Phys. Rev. {\bf A86}, 063635 (2012).

% 1d superlattice
\bibitem{buonsante04} P. Buonsante, V. Penna, A. Vezzani,
Phys. Rev {\bf A70}, 061603R (2004).

\bibitem{buonsante05} P. Buonsante, A. Vezzani, Phys. Rev. {\bf A72},
  013614 (2005).

\bibitem{rousseau06}
%% Exact Study of the $1D$ Boson
%% Hubbard Model with a Superlattice Potential,
V.G.~Rousseau, D.P.~Arovas, M.~Rigol, F.~H\'ebert,
G.G.~Batrouni, and R.T.~Scalettar,
Phys. Rev. {\bf B73}, 174516 (2006).

% 2d superlattice

\bibitem{buonsante05b}  P. Buonsante, V. Penna, and A. Vezzani, Phys.
Rev. {\bf A72}, 031602(R) (2005).

\bibitem{santos04} L. Santos, M. A. Baranov, J. I. Cirac,
  H.-U. Everts, H. Fehrmann and M. Lewenstein, Phys. Rev. Lett. {\bf
    93}, 030601 (2004).

%superlattice chains
\bibitem{danshita07} I. Danshita, J. E. Williams, C. A. R. S\'a de Melo
  and C. W. Clark, Phys. Rev. {\bf A76}, 043606 (2007).

\bibitem{danshita08} I. Danshita, C. A. R. S\'a de Melo and C. W. Clark,
  Phys. Rev. {\bf A77}, 063609 (2008).



\bibitem{girardeau60}
M. Girardeau, J. Math. Phys. {\bf 1}, 516 (1960).

\bibitem{jordan28}
P. Jordan and E. Wigner, Z. Phys. {\bf 47}, 631 (1928).

\bibitem{crepin11}
F. Crepin, N. Laflorencie, G. Roux, and P. Simon,
Phys. Rev. {\bf B 84}, 054517 (2011).

\bibitem{alps11} B. Bauer, L. D. Carr, H.G. Evertz, A. Feiguin,
  J. Freire, S. Fuchs, L.  Gamper, J. Gukelberger, E. Gull,
  S. Guertler, A. Hehn, R. Igarashi, S.V.  Isakov, D. Koop, P.N. Ma,
  P. Mates, H. Matsuo, O. Parcollet, G.  Pawlowski, J.D. Picon,
  L. Pollet, E. Santos, V.W. Scarola, U.  Schollw\"ock, C. Silva,
  B. Surer, S. Todo, S. Trebst, M. Troyer, M.L.  Wall, P. Werner, and
  S. Wessel, J. Stat. Mech. (2011) P05001.

\bibitem{sandvik99}
A. W. Sandvik, Phys. Rev. {\bf B 59}, 14157 (1999).

\bibitem{alet05} F. Alet, S. Wessel, and M. Troyer, Phys. Rev. {\bf E
    71}, 036706 (2005).

\bibitem{pollet05}
L. Pollet, S. M. A. Rombouts, K. Van Houcke, and K. Heyde,
Phys. Rev. {\bf E 70}, 056705 (2005).

\bibitem{handscomb62}
D.C. Handscomb, Proc. Cambridge Philos. Soc. {\bf 58}, 594 (1962).

\bibitem{pollock} E.L. Pollock and D.M. Ceperley, Phys. Rev. {\bf B30},
2555 (1984);
D.M. Ceperley and E.L. Pollock, Phys. Rev. Lett. {\bf 56},
351 (1986); and
E.L. Pollock and D.M. Ceperley, Phys. Rev. {\bf B36},
8343 (1987).

\bibitem{ying13}
T. Ying, G.G. Batrouni, V.G. Rousseau, M. Jarrell,
J. Moreno, X.D. Sun, and R.T. Scalettar,
Phys. Rev. {\bf B 87}, 195142 (2013).

\end{thebibliography}
\end{document}